\documentstyle[epsf,wrapfig]{ptptex}
\def\Journal#1#2#3#4{{#1} {\bf #2} (#4) #3}
\def\PTP{\em Prog. Theor. Phys.}

\def\PLB{{\em Phys. Lett.}  B}
\def\PL{{\em Phys. Lett.} }
\def\PRL{\em Phys. Rev. Lett.}

\def\PRD{{\em Phys. Rev.} D}
\def\PR{{\em Phys. Rev.} }
\def\ZPC{{\em Z. Phys.} C}

\title{%
Spectra of Exclusive Semi-Leptonic Decays of B-meson\\
in the Covariant Oscillator Quark Model
}
\author{%
Muneyuki {\sc Ishida}, Shin {\sc Ishida}${}^a$ and Masuho {\sc Oda}${}^b$
}
\inst{%
Department of Physics, University of Tokyo , Tokyo 113\\
${}^a$Atomic Energy Research Institute, College of Science and\\ 
\hspace*{2.8cm}Technology, Nihon University, Tokyo 101\\
${}^b$Faculty of Engineering, Kokushikan University, Tokyo 154 
}
\recdate{%
February 11, 1997
}
\abst{%
The spectra and form factors of exclusive semi-leptonic
decays of B-meson are analyzed, taking account of confined effects
of quarks by the Covariant Oscillator Quark Model. 
The predicted $B\rightarrow (D^*,D)l\bar{\nu}_l$ 
spectra  with the conventional value of $V_{cb}$
are fit well to the high-precision experiment.
It is also shown the values of $V_{cb}$ and $V_{ub}$
obtained through decay widths for the 
$B\rightarrow (D,D^*,\rho )l\bar{\nu}_l$ processes are consistent
with the presently accepted ones.
}

\begin{document}
\maketitle

\section{Introduction}
The exclusive semi-leptonic decay has been one of the important topics of 
high energy physics for many years, however it has been 
a difficult task to explain
its spectra quantitatively. They are largely affected by
the confined effects of quarks.
Especially in the decays of B-mesons the final mesons
generally make highly relativistic motions, making the 
situation more complex.
Among many experimental studies
the spectra of $B\rightarrow D^*l\bar{\nu}_l$ 
and $B\rightarrow Dl\bar{\nu}_l$ decays are 
obtained with high precisions,
of which analysis has been one of the main interests in the
heavy quark effective theory (HQET)\cite{rf:neubert}.
In HQET all generally independent form factors, 
appearing in the effective meson transition 
current $J_\mu^{B\rightarrow D^*/D}$, 
are represented by one universal form factor(FF) function
(Isgur-Wise function $\xi (\omega )$), thus leading to various FF-relations 
among them. 
The value of $\xi (\omega )$ at zero-recoil point
is to be unity, $\xi (1)=1$, in the heavy quark mass limit
$m_Q\rightarrow\infty$, reflecting
the conserved charge of heavy quark symmetry (HQS).
However, HQET and/or HQS  themselves are not able to
predict the concrete form of $\xi (\omega )$, and accordingly
it cannot describe the FF function and 
the decay spectra in all regions of $q^2$.
For this it is, in principle, 
necessary to know covariant wave functions(WF)
of mesons concerning both spin and space-time variables,
and presently we are required to resort 
to some models with a covariant framework.
The FF function is obtained as an overlapping of initial and 
final state WF.

In this work we shall study the FF's and spectra of 
semi-leptonic  B-meson decays,
using the Covariant Oscillator Quark Model (COQM)
in order to estimate the confined effects of quarks.
The
COQM has a long history of development\cite{rf:COQM},
and its origin may be traced back to the bilocal theory by Yukawa.\cite{rf:yukawa}
The general framework of COQM is called the boosted LS-coupling scheme,
and the hadron WF, being tensors\cite{rf:spin} in 
$\tilde{U}(4)\times O(3,1)$-space,
reduce
to those in $SU(2)_{\rm spin}\times O(3)_{\rm orbit}$-space
in the non-relativistic quark model in the hadron rest frame. 
The spinor and space-time portions of 
WF satisfy separately the respective covariant equations, the
Bargmann-Wigner(BW) 
equations for the former and the covariant oscillator equation
for the latter. 
The concrete form of meson
WF has been determined 
completely through the analysis of 
mass spectra\cite{rf:mesonmass,rf:yamada}.
Their validity seems to be shown, to some extent,
phenomenologically by their applications to
various dynamical processes, such as the
electro-magnetic FF\cite{rf:kobayashi}
and the radiative decay of light quark mesons.\cite{rf:radiative} 
Recently we have applied them to
general 
semi-leptonic decays\footnote{
Application  of covariant oscillator function 
to semi-leptonic decays was done firstly in ref.\cite{rf:kizukuri}. 
The preliminary result of this work was presented in
refs.\cite{rf:oda}.
}
of mesons and baryons.
It is especially interesting that for the $B\rightarrow D^*/Dl\bar{\nu}_l$ 
decays  completely 
the same\cite{rf:oda,rf:effective} 
FF relations as in HQET are derived in COQM.
The physical origin of obtaining the same FF 
relations is the use of the BW spinor functions.
We have argued in our previous work that 
these BW functions are also implicitly supposed\cite{rf:spin,rf:oda}
in HQET. The reduced FF function in COQM, 
called $I$-function\footnote{
In the case of light-and heavy-quark meson system the $I$-function
(taking $m_Q\rightarrow\infty$) is identical to 
the Isgur-Wise function $\xi (\omega )$.
}, is shown to
satisfy also the above mentioned condition of the $\xi (1)=1$.

According to the analysis\cite{rf:spin,rf:mesonmass,rf:yamada}
of meson mass spectra, there seems to be a 
sufficient phenomenological reason for the 
validity of BW equations also in the light quark meson system.
So we shall make the similar analysis of the process
$B\rightarrow \rho l\bar{\nu}_l$\cite{rf:oda},
(which is out of the scope of HQET by itself) as well as 
$B\rightarrow D/D^*l\bar{\nu}_l$ in this work.

\section{Meson wave functions}
In COQM all non-exotic $q\bar{q}$-mesons are described unifiedly by 
bi-local fields
${\Phi_A}^B(x_{1\mu},x_{2\mu})$,
where $x_{1\mu}(x_{2\mu})$ is a space-time coordinate of constituent quark(anti-quark),
$A=(a,\alpha )\ (B=(b,\beta ))$ describing its flavor and
covariant spinor.
Here we write only the (positive frequency part of)
relevant ground state fields.\cite{rf:spin}
\begin{eqnarray}
{\Phi_A}^B(x_{1\mu},x_{2\mu})
 &=& e^{iP\cdot X}U(P{)_A}^B {f_{(ab)}}(x_\mu ;P),
\label{eq:wf}
\end{eqnarray}
where $U$ and $f$ are the covariant spinor 
and internal space-time WF,
respectively, satisfying the BW and oscillator wave equations.
The $x_\mu$
$(X_\mu)$
is the relative (CM) coordinate,
$x_\mu\equiv x_{1\mu}-x_{2\mu}$
$(X_\mu\equiv (m_1x_{1\mu}+m_2x_{2\mu})/(m_1+m_2) $;
 $m_i$'s being quark masses).
The $U$ is given by
\begin{eqnarray}
U(P)
 &=& \frac{1}{2\sqrt{2}}[(-\gamma_5P_s(v)+i\gamma_\mu V_\mu (v))
(1+iv\cdot\gamma)],
\label{eq:swf}
\end{eqnarray}
where $P_s(V_\mu )$ represents the pseudoscalar (vector)
meson field, and $v_\mu \equiv P_\mu /M(P_\mu (M)$ being 
four momentum(mass) of meson). 
The $U$, being represented by the direct product of
quark and anti-quark Dirac spinor with the
meson velocity, is reduced to
the non-relativistic Pauli-spin function in the meson rest frame.
The $f$ is given by
\begin{eqnarray}
f(x_\mu ;P)
 &=& \frac{\beta}{\pi}e^{-\frac{\beta}{2}\left( 
    x_\mu^2+2\frac{(x\cdot P)^2}{M^2} \right)    },\ \ \ \ 
\beta =\sqrt{\mu K},
\label{eq:stwf}
\end{eqnarray}
where $\mu$ is the reduced mass, and $K$ is the spring constant
of oscillator potential.

We treat our problem in the following two cases of parameters,
$m_q'$s and $K$.

Case A: The $m_q'$s are simply determined by
$M_V=m_q+m_{\bar{q}}$, $M_V$'s being the mass of 
relevant vector mesons, $B^*,\ D^*$ and $\rho$.
The $K$ is supposed to be universal\cite{rf:mesonmass}, 
independently of flavor-contents
of mesons, and determined from 
the Regge slope of $\rho$ meson trajectory,
$\Omega^{\rm exp}=1.14$ GeV, by a relation
 $\Omega =\sqrt{32m_nK}$.

Case B: They are determined from the recent analyses 
of mass spectra, including the effects due to 
color Coulomb force\cite{rf:yamada}.
In this case the mixing of ground 1S-states with excited 2S-states
is shown to be a few percents (the less)
for $\rho$ ($D$ or $D^*$ and $B$) in the amplitudes,
and thus its effects seem to be negligible.

The actual values of $m_q$'s and the $\beta$ for the respective systems 
are collected in Table I. The ``size'' of wave functions $\beta^{-1}$ 
seem to be almost
equal in case B. 
\begin{table}
\caption{The adopted values of quark masses  $m_q$'s and 
of inverse sizes $\beta$'s.  
In Case A:  
$K=0.106$ GeV${}^3$ is taken to be universal.  
In Case B: $K$ is determined from mass
spectra as $K=$
0.0979, 0.0679 and 0.0619 GeV${}^3$, respectively,
for $\rho$, $D(D^*)$ and $B$ mesons.
In Case B the size of mesons becomes almost equal.}
\begin{center}
\begin{tabular}{|l|c|c|c||c|c|c|}
\hline
    & $m_n$ & $m_c$ & $m_b$ & $\beta_\rho$ & $\beta_{D/D^*}$ & $\beta_B$ \\
\hline
Case A & 0.384 GeV & 1.62 GeV & 4.94 GeV & 0.143 GeV${}^2$ 
                                &  0.181 GeV${}^2$ & 0.194 GeV${}^2$ \\
\hline
Case B & 0.400 GeV & 1.70 GeV & 5.00 GeV & 0.140 GeV${}^2$ 
                                &  0.148 GeV${}^2$ & 0.151 GeV${}^2$ \\
\hline
\end{tabular}
\end{center}
\end{table}

\section{Effective weak currents and reduced form factor function}
Our effective action for weak interactions of mesons with W-boson is given
by
\begin{eqnarray}
S_W &=& \int d^4x_1d^4x_2 \langle\bar{\Phi}_{F,P'}(x_1,x_2)i\gamma_\mu
(1+\gamma_5 )\Phi_{I,P}(x_1,x_2)\rangle W_{\mu ,q}(x_1),
\label{eq:mech}
\end{eqnarray}

\begin{wrapfigure}{r}{6.cm}
  \epsfysize=4.5cm
 \centerline{\epsffile{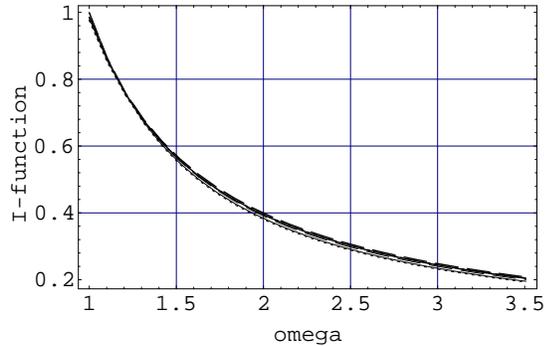}}
  \caption{$I$-functions of  $B\rightarrow D^*$, $B\rightarrow D$
           and $B\rightarrow \rho$ transitions, are denoted by
           solid, dashed and dotted lines, respectively.
           The thick (thin) line corresponds to the case A(B). }
  \label{fig:one}
\end{wrapfigure}
\hspace*{-0.7cm}where we have denoted the interacting (spectating) quarks as 1(2).
(the KM matrix elements and the coupling constant are omitted.)
This is obtained  from consideration of Lorentz covariance, supposing
the quark current with the standard $V-A$ form.
In Eq.(\ref{eq:mech})
$\Phi_{I,P}\ (\bar{\Phi}_{F,P'})$ denotes the initial (final) meson
with the definite four momentum $P_\mu\ (P_\mu ')$,
and $q_\mu$ is the momentum of W-boson.
The $\bar{\Phi}$ is the Pauli-conjugate of $\Phi$ defined by
$\bar{\Phi} \equiv - \gamma_4 \Phi^{\dag} \gamma_4$ and $\langle\ \ \rangle$
represents the trace of Dirac spinor indices.
Our relevant effective currents $J_\mu (X)_{P',P}$ 
are obtained by identifying 
the above action with
\begin{eqnarray}
S_W &=& \int d^4X  J_\mu (X)_{P',P}W_\mu (X)_q.
\label{eq:Jmu}
\end{eqnarray}
Then $J_\mu (X=0)_{P',P}\equiv J_\mu$ is explicitly 
given\cite{rf:spin,rf:effective} as
\begin{eqnarray}
J_\mu &=& I_n^{qb}(\omega )\sqrt{MM'} \nonumber\\
 & & [\bar{P}_s(v')P_s(v)(v+v')_\mu 
  + \bar{V}_\lambda (v')P_s(v)
(\epsilon_{\mu\lambda\alpha\beta}v'_\alpha v_\beta -\delta_{\lambda\mu}
(\omega +1)-v_\lambda v'_\mu )],
\label{eq:current}
\end{eqnarray}
where $\omega =-v\cdot v'$.
The $I_n^{qb}(\omega )$ is the overlapping of
the initial and final space-time wave functions, 
which describes the confined effects of quarks, and is given by
\begin{eqnarray}
I_n^{qb}(\omega ) &=& 
\frac{4\beta\beta '}{\beta +\beta '}\frac{1}{\sqrt{C(\omega )}} 
exp(-G(\omega ));
\ \ \ C(\omega )=(\beta -\beta ')^2+4\beta\beta '\omega^2
\label{eq:Ifunc}\\
G(\omega ) &=& \frac{m_n^2}{2C(\omega )}\left[
(\beta+\beta ')
   \left( 
          \left(\frac{M}{M_{\rm s}}\right)^2
         + \left(\frac{M'}{M_{\rm s}'}\right)^2
         -2\frac{MM'}{M_{\rm s}M_{\rm s}'}\omega 
   \right) \right. \nonumber\\
&\ \ & +2
   \left.
   \left( 
          \beta '\left(\frac{M}{M_{\rm s}}\right)^2
         +\beta  \left(\frac{M'}{M_{\rm s}'}\right)^2 
   \right)(\omega^2-1)
\right] ,
\label{eq:Gfunc}
\end{eqnarray}

\begin{wrapfigure}{r}{5.5cm}
  \epsfysize=7cm
  \centerline{\epsffile{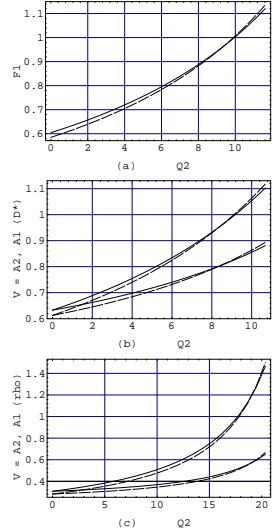}}
  \caption{ $Q^2$-dependence of transition form factors: 
(a) $F_1$ in $B\rightarrow D$,
(b) $V=A_2$ (thin line) and $A_1$(thick line)
       in $B\rightarrow D^*$,
(c) $V=A_2$ (thin line) and $A_1$(thick line)
       in $B\rightarrow \rho$.
     The solid (dashed) line corresponds to case A(B)
. }
  \label{fig:two}
\end{wrapfigure}
\hspace*{-0.7cm}where $M_{\rm s}(M_{\rm s}')$ is the sum of the constituent quark masses,
$m_1+m_2(m_1'+m_2')$,
of the initial (final) meson system. 
The actual forms of relevant I-functions 
are drawn in Fig. 1, which
are almost common
in $B\rightarrow D^*,D$ and $\rho$ transitions.
However, the kinematically allowed 
maximum value $\omega_{\rm max}$ of $\omega$, 3.51, in $B\rightarrow\rho$
transition is much larger than those 1.50(1.59)
in $B\rightarrow D^*$ ($B\rightarrow D$), 
and the corresponding value of $I(\omega_{\rm max})$
is much smaller for the former than for the latter.

If we identify $M(M')$ with $M_s(M_s')$,
the $G(\omega )$, Eq.(\ref{eq:Gfunc}), becomes simple as
$G(\omega)=m_n^2(\beta +\beta ')\omega (\omega -1)/C(\omega ).$
Moreover,
in the $m_Q\rightarrow\infty$ limit 
the size of $B$ and $D(D^*)$ mesons becomes equal, 
that is, $\beta_B=\beta_{D/D^*}\equiv\beta_\infty$.
Then the $I$-function of $B\rightarrow D/D^*$ transitions 
takes the form, 
$I_\infty  (=\xi (\omega ))= 1/\omega\  
exp[ -m_n^2(\omega -1)/(2\beta_\infty \omega )],$
which has the property at zero-recoil, 
$I_\infty (\omega =1 )=1$,
similarly
as the $\xi (\omega )$.
The actual values of $I(1)$ 
in $B\rightarrow D$ and $B\rightarrow D^*$ transitions are, respectively,
0.984(0.999) and 0.986(1.000) for case A(B).

\section{Form factors and Decay spectra}

The invariant FF's of $B\rightarrow D^*/D$ transitions 
are defined as:
\begin{eqnarray}
\langle D^*(P',\epsilon )|J_\mu |B(P)\rangle
 &=&   \frac{2\epsilon_{\mu\lambda\alpha\beta}}{M+m_{D^*}}
       \tilde{\epsilon}_\lambda  
           P'_\alpha P_\beta V^{D^*}(q^2) 
 - (M+m_{D^*})\tilde{\epsilon}_\mu 
A_1^{D^*}(q^2) \nonumber\\ 
 - \frac{\tilde{\epsilon}\cdot q}{M+m_{D^*}}(P+P')_\mu 
 & A_2^{D^*} &, \ \ \ \    
\langle D(P')|J_\mu |B(P)\rangle = F_1^D(q^2)(P+P')_\mu ,
\label{eq:ffdef}
\end{eqnarray}
neglecting the terms proportional to $q_\mu$, which are irrelevant
in the massless lepton approximation.
Comparing Eq.(\ref{eq:ffdef}) with 
our effective currents Eq.(\ref{eq:current}) leads to the 
FF's represented by $I$-function as
\begin{eqnarray}
V^{D^*} = A_2^{D^*} = \frac{A_1^{D^*}}{1+q^2/(M+m_{D^*})^2}
 =  \frac{M+m_{D^*}}{2\sqrt{Mm_{D^*}}}  I_n^{cb};\ \ 
F_1^D = \frac{M+m_D}{2\sqrt{Mm_D}}I_n^{cb}.\ \ \ \ 
\label{eq:ffrel}
\end{eqnarray}
In $B\rightarrow\rho$ transition the FF's, defined similarly 
as in Eq.(\ref{eq:ffdef}), are represented by
\begin{eqnarray}
V^{\rho} &=& A_2^{\rho}
 =   \frac{A_1^{\rho}}{1+q^2/(M+m_{\rho})^2}=
\frac{M+m_{\rho}}{2\sqrt{Mm_{\rho}}}I_n^{ub}(\omega ).
\label{eq:rhorel}
\end{eqnarray}
These  FF relations are the same as in
$B\rightarrow D^*$ and $B\rightarrow\rho$\footnote{
A similar formulus for $B\rightarrow\pi$ transition as for $B\rightarrow D$
may be also applicable.
However, it may not be effective because no considerations on 
the property of $\pi$
as Nambu-Goldstone boson are given in our present framework.} 
transitions, according to common
use of BW spinor function.
The actual $Q^2(\equiv -q^2)$ dependence of 
relevant FF's are given in Fig. 2.

Our theoretical spectra of
$B\rightarrow Dl\nu_l$ and
$B\rightarrow Vl\nu_l\ (V=D^*,\ \rho )$ decays 
are obtained from the effective currents (\ref{eq:current}) as
\begin{eqnarray}
\frac{d\Gamma^{B\rightarrow D}}{dq^2} &=& 
 |V_{cb}|^2(I_n^{cb})^2\frac{G_F^2m_D^2}{96\pi^3M_B}\sqrt{\omega^2-1}
(\omega^2-1)(M_B+m_D)^2, \\
\frac{d\Gamma^{B\rightarrow V}}{dq^2} &=& 
 |V_{qb}|^2(I_n^{qb})^2\frac{G_F^2m_V^2}{96\pi^3M_B}\sqrt{\omega^2-1}
(\omega +1)[-4\omega q^2+(\omega +1)(M_B-m_V)^2], \nonumber
\label{eq:spectra}
\end{eqnarray}
where $V_{qb}$'s are the relevant Kobayashi-Maskawa matrix 
elements $V_{\rm KM}$ (, our only free parameters,) and $G_F$
is the Fermi constant (we adopt the value $1.166\times 10^{-5}$GeV${}^{-2}$). 
The best fits to the experimental spectra,
(a) $|V_{cb}|I(\omega )$ and (b) $d\Gamma/dQ^2$, are shown,
in Fig.3 for $B\rightarrow D^*$\cite{rf:cleo} decay and 
in Fig.4 for $B\rightarrow D$\cite{rf:cleonew} decay. 
For reference we have also shown the results
with $I$-function taken as unity,
which correspond to ``free quark decay,"\footnote{
Here the combined value of $V_{cb}$ in Table II
is used.
} in  Figs. 3(b) and 4(b).
The $V_{cb}$-values thus determined are given in Table II,
where the combined values of $V_{cb}$ for the respective processes
are also given.
\begin{table}
\caption{The values of $V_{cb}$ obtained from the best fit to the experimental 
spectra in $B\rightarrow D^*l\bar{\nu}_l^{12)}$ 
and $B\rightarrow Dl\bar{\nu}_l^{13)}$ decays. 
The second errors are the systematic ones
for $|V_{cb}|F(1)$(including the error of B-life time)
quoted from ref.12) and ref.13) for the respective processes.   }
\begin{center}
\begin{tabular}{|l|c|c|c|}
\hline
          & case A & case B & combined\\
\hline
 $B\rightarrow D^*l\bar{\nu}_l$     & $(3.88\pm 0.10\pm 0.20)\times 10^{-2}$ 
                                    & $(3.89\pm 0.10\pm 0.20)\times 10^{-2}$
                                    & 
$(3.89\pm 0.10\pm 0.20)\times 10^{-2}$ \\
\hline
 $B\rightarrow Dl\bar{\nu}_l$     & 
$(4.11\pm 0.14\pm 0.46)\times 10^{-2}$ 
                                  & 
$(4.15\pm 0.14\pm 0.46)\times 10^{-2}$ 
                                  & 
$(4.13\pm 0.14\pm 0.46)\times 10^{-2}$  \\
\hline
\end{tabular}
\end{center}
\end{table}

\begin{wrapfigure}{c}{16cm}
  \epsfysize=5cm
  \centerline{\epsffile{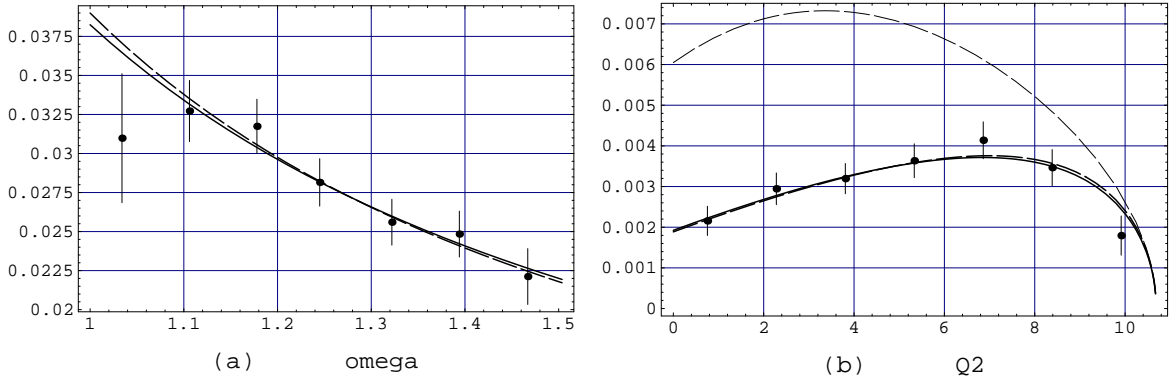}}
  \caption{Comparison of theory and experiment${}^{12)}$ for decay spectra 
         in $B\rightarrow D^*l\bar{\nu}_l$, 
   (a) $|V_{cb}|I(\omega )$ and (b)$d\Gamma/dQ^2$(ps${}^{-1}$GeV${}^{-2}$). 
         A thick solid (dashed) line corresponds to the case A(B). 
   A thin dashed line in Fig.(b) is the one
   in case of ``free quark decay".    
  }
  \label{fig:three}
\end{wrapfigure}
\begin{wrapfigure}{c}{16cm}
  \epsfysize=5cm
  \centerline{\epsffile{figfournew.epsi}}
  \caption{Comparison of theory and experiment${}^{13)}$ for decay spectra 
         in $B\rightarrow Dl\bar{\nu}_l$, 
  (a) $|V_{cb}|I(\omega )$ and (b)$d\Gamma/dQ^2$(ps${}^{-1}$GeV${}^{-2}$). 
         A thick solid (dashed) line corresponds to the case A(B). 
   A thin dashed line in Fig.(b) is the one
   in case of ``free quark decay".    
  }
  \label{fig:fournew}
\end{wrapfigure}
As is seen in Figs. 3(b) and 4(b), the confined effect is small in the non-relativistic region
with large $Q^2\ (\omega\sim 1)$, and the predicted spectra 
are close to the ones in the case of free quark decay.
In the relativistic region
with small $Q^2\ (\omega\sim\omega_{\rm max})$
the confined effect becomes very large.
It is interesting that we are led to 
good fits as a result of this large effects.

As is seen from Table II
the values of $V_{cb}$ are mutually consistent for $B\rightarrow D^*$
and $B\rightarrow D$ decays.
This is required from the viewpoints of HQET and  
the boosted LS-coupling scheme
 
The predicted spectrum for $B\rightarrow \rho l\nu_l$ decay 
in unit of $V_{\rm ub}$, 
$d\tilde{\Gamma}/dq^2\equiv d\Gamma /(dq^2|V_{\rm ub}|^2)$,
is shown in 
Fig. 5, where the spectrum for ``free quark decay" is also given.
It is seen that 
the confined effects
drastically change the spectrum of $B\rightarrow\rho$ decay.

In Fig. 6 the decay spectra for $B\rightarrow D^*l\bar{\nu}_l$ and 
$B\rightarrow\rho l\bar{\nu}_l$  
predicted by COQM are shown and compared 
with the other models, WSB\cite{rf:wsb}, KS\cite{rf:ks}, ISGW\cite{rf:isgw} 
and ISGW2\cite{rf:isgw2} models.

\begin{wrapfigure}{r}{5cm}
  \epsfysize=4cm
  \centerline{\epsffile{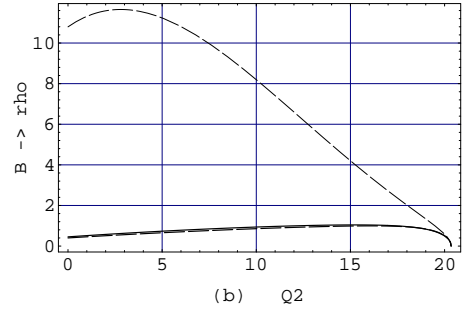}}
  \caption{ Predicted
spectrum of $B\rightarrow \rho l\nu_l$ decay. 
 A thick solid (dashed) line 
  corresponds to the case A(B). The spectrum for
``free quark decay" is shown with
thin dashed line.}
  \label{fig:four}
\end{wrapfigure}
All models give similar results for 
$B\rightarrow D^*l\bar{\nu}_l$, while much different ones for
$B\rightarrow\rho l\bar{\nu}_l$. 
The spectra for the latter predicted by COQM are similar to 
those of
WSB, KS (ISGW) model in the relativistic (non-relativistic)
region of $Q^2$, that is, $Q^2\sim Q^2_{\rm max}$ ($Q^2\sim 0$).
This fact seems to reflect the theoretical backgrounds of the 
respective models.
In WSB and KS models, the absolute values of FF's at $Q^2=0$ are fixed by the 
overlapping of wave functions of the scalar harmonic oscillator
model in the infinite momentum frame. 
The ISGW model is essentially a non-relativistic (NR) model,
and the FF's are given by the overlapping of NR-meson wave functions,
which are determined through the analysis of mass spectra of meson systems.
Accordingly it may be reliable in the NR-region, $Q^2\sim Q^2_{\rm max}$.
The spectra of ISGW2 model, which is updated from the original model, 
become closer in the relativistic region to those of COQM,
which has a reliable kinematical framework in both relativistic and non-relativistic regions. 

\begin{wrapfigure}{c}{16cm}
  \epsfysize=5.5cm
  \centerline{\epsffile{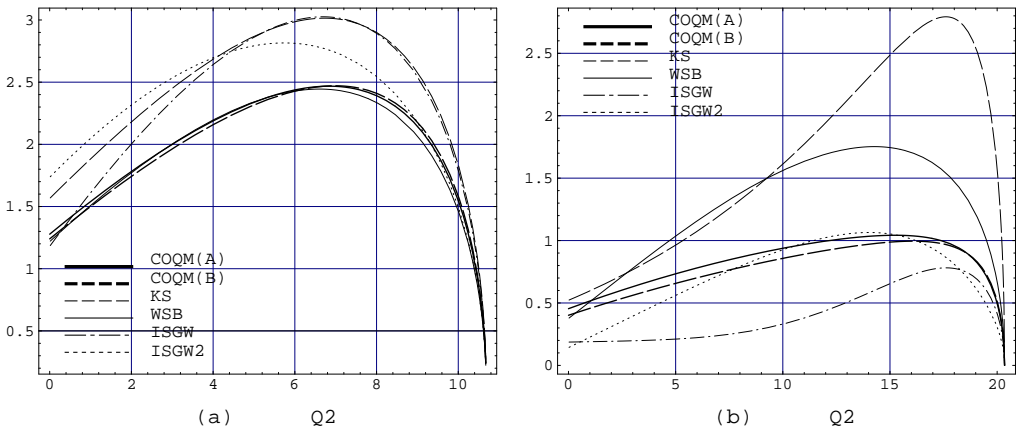}}
\caption{
 Comparison of our spectra  
 $d\tilde{\Gamma}/dQ^2\equiv d\Gamma/dQ^2|V_{qb}|^2$
 with the other models:
 (a) $B\rightarrow D^* l\nu_l$  and 
 (b) $B\rightarrow \rho l\nu_l$ decays. 
  Thick solid (dashed) line 
  corresponds to the case A(B) of COQM. Respective thin lines
  correspond to the models of KS(dashed), 
  WSB(solid), ISGW(dot-dashed) and ISGW2(dotted).
}
  \label{fig:five}
\end{wrapfigure}
Presently there have been reported no experimental spectra 
for the $B\rightarrow \rho l\bar{\nu}_l$ decay. 
However,
by comparing the theoretical values of
$\tilde{\Gamma}(\equiv\Gamma /|V_{\rm KM}|^2)$ with
the experimental decay widths $\Gamma '$s for 
$B\rightarrow D/D^* l\bar{\nu}_l$ and
$B\rightarrow \rho l\bar{\nu}_l$ processes,
we can also determine the values of $V_{\rm KM}$-elements.
The relevant values of $V_{\rm KM}$ thus obtained 
are collected and compared with the results 
obtained in the above models in Table III. There we have also 
given the results in the case of free quark decay.
It is to be noted that
$\tilde{\Gamma}^{\rm theor}$'s
are much smaller, especially in $B\rightarrow \rho$ decay, 
than $\tilde{\Gamma}_{\rm free}^{\rm theor}$'s.
The $V_{cb}$-values in Table III are consistent to those in Table II
obtained from the decay spectra.
They are nearly the same as the estimate by
HQET, 0.038$\pm$0.007\cite{rf:neubert}. Our $V_{ub}$-value
is also consistent to the conventional estimate\cite{rf:PDG96},
$(2\sim 5)\times 10^{-3}$, derived
by various other methods. 
From Table III we see that especially
the $\tilde{\Gamma}$ for $B\rightarrow \rho l\bar{\nu}_l$ shows large
model-dependence, and experimental studies on this process
seem important to select the models.


\begin{table}
\caption{ Our values of $V_{cb}$ and $V_{ub}$ estimated from decay widths
of $B\rightarrow D$
$B\rightarrow D^*$
and $B\rightarrow\rho$
processes. The theoretical values without(with) 
brackets corresponds to 
case A(B). $\tilde{\Gamma}^{\rm theor}_{\rm free\ q}$'s are 
the widths in the case of ``free quark 
decay". $\tilde{\Gamma}$'s obtained in the other models 
(WSB, KS, ISGW and ISGW2) are also given.}
\begin{center}
\begin{tabular}{|l|c|c|c|}
\hline
    & $B\rightarrow Dl\nu_l$ & $B\rightarrow D^*l\nu_l$
    & $B\rightarrow \rho l\nu_l$  \\
\hline
 $\Gamma^{\rm exp}$ 
         & \cite{rf:cleonew}$(11.3\pm 2.0)\times 10^{-3}$ ps${}^{-1}$
         & \cite{rf:cleo}$(29.9\pm 3.9)\times 10^{-3}$ ps${}^{-1}$
         & \cite{rf:moneti}$(16.0\pm 5.6)\times 10^{-5}$ ps${}^{-1}$\\
\hline
 $V_{\rm KM}$ & $V_{cb}$=0.0396$\pm$0.0036 
          & $V_{cb}$=0.0372$\pm$0.0025 
        & $V_{ub}=(3.1\pm 0.5)\times 10^{-3}$ \\
          & \ \ \ \ \ (0.0399$\pm$0.0036)
          & \ \ \ \ \ (0.0374$\pm$0.0025)
          & \ \ \ \ \ $((3.2\pm 0.6)\times 10^{-3}$)\\
\hline
 $\tilde{\Gamma}^{\rm theor}_{\rm free\ q}$ 
          & 18.7 ps${}^{-1}$ & 41.9 ps${}^{-1}$ & 150 ps${}^{-1}$ \\
\hline
 $\tilde{\Gamma}^{\rm theor}$ 
         & 7.18(7.06) ps${}^{-1}$ & 21.6(21.4) ps${}^{-1}$ 
         & 17.0(15.8) ps${}^{-1}$ \\
\hline
\hline
$\tilde{\Gamma}$(WSB) & 8.08 ps${}^{-1}$ & 21.9 ps${}^{-1}$ & 26.1 ps${}^{-1}$ \\
\hline
$\tilde{\Gamma}$(KS) & 8.3 ps${}^{-1}$ & 25.8 ps${}^{-1}$ &  33.0 ps${}^{-1}$ \\
\hline
$\tilde{\Gamma}$(ISGW) & 11 ps${}^{-1}$ & 25 ps${}^{-1}$ &  8.3 ps${}^{-1}$ \\
\hline
$\tilde{\Gamma}$(ISGW2) & 11.9 ps${}^{-1}$ & 24.8 ps${}^{-1}$ &  14.2 ps${}^{-1}$ \\
\hline
\end{tabular}
\end{center}
\end{table}

\section{Concluding Remarks}
We have studied, applying COQM, the spectra
of semi-leptonic decay of B-meson, $B\rightarrow D^* l\bar{\nu}_l$
and $B\rightarrow D l\bar{\nu}_l$,
which are known experimentally with high precisions, 
and obtained the consistent results to
the experiments as a result of taking account of large 
quark confined effects.
The values of $V_{cb}$ determined from these two decay spectra 
are mutually consistent.
The $B\rightarrow D,D^*$ and $\rho$ $l\bar{\nu}_l$ decay widths are 
also reproduced satisfactorily with the presently accepted values of the relevant
$V_{\rm KM}$-elements. These facts seem to give an evidence for validity of the
framework of COQM, the boosted LS-coupling scheme,
describing the meson systems.

The framework of COQM is also applicable directly to the transitions 
between hadrons including 
excited states, such as $B\rightarrow D^{**}l\bar{\nu}_l$,
which will be treated elsewhere.\\
Finally we should like to express our gratitude to Dr. K.Yamada,
for informing us of the recent results of studies on mass spectra.

\end{document}